\documentstyle[11pt,epsfig,epsf]{article}
\newcommand{\beq}{\begin{equation}}
\newcommand{\eneq}{\end{equation}}

\begin{document}


\hfill{Napoli Preprint DSF-2000-34}
\large
\baselineskip=16pt

{\Large \bf
\begin{center}
{The global phase diagram of a modular invariant two dimensional
statistical model}
            \footnote{Work supported in part by the European Commission RTN
programme HPRN-CT-2000-00131 .}
\end{center} }

\begin{center}
G. Cristofano$^{* \dagger}$, D. Giuliano$^{\dagger \dagger}$, G.
Maiella$^{*
\dagger}$

 \vspace{0.2cm}

{\normalsize $^{*}$ Dipartimento di Scienze Fisiche, Universit\`{a}  di Napoli
\\
$^{\dagger}$ INFN - Sezione di Napoli \\ Via Cintia - Compl.\ universitario
M. Sant'Angelo
- 80126 Napoli, Italy

$^{\dagger \dagger}$ Department of Physics, Stanford University, Stanford,
California 94305}
\end{center}

\vspace{1cm}

\begin{center}
\end{center}

\begin{center}
\end{center}

\vspace{1cm}

{\small
\begin{quote} {\bf Abstract} A generalization of the Coulomb Gas model
with modular $SL(2, Z)$-symmetry
 allows for a discrete infinity of phases which are characterized by the condensation of
 dyonic pseudoparticles and the breaking of parity and time reversal.
 Here we study the phase diagram of such a model by using renormalization
group techniques. Then the symmetry $SL(2,Z)$ acting on the two-dimensional
parameter space gives us a nested shape of its global phase diagram and all
the infrared stable fixed points. Finally we propose a connection with the
2-dimensional Conformal Field Theory description of the Fractional Quantum
Hall Effect.
\end{quote}}


\setcounter{page}{1}
\newpage

\section{Introduction.}

In 1982 Cardy and Rabinovici, following an original suggestion by G.
$~^{'}$t Hooft \cite{gerard}, have built a simple model derived by means of
dimensional reduction of an Abelian $Z_p$ lattice gauge  theory in four
dimensions \cite{car1,car2}. The relevant parameters of the theory are the
coupling strength $g$ and the $\theta$-angle. By integrating out the gauge
fields and by using the electric and magnetic current densities as
dynamical variables, the partition function of such a model is mapped onto
the one of a ``generalized" or ``modular invariant Coulomb Gas" ({\bf
mCG}). The model does preserve both parity ({\bf P}) and time-reversal
({\bf T}) symmetries and so does any of its phases, given by condensates
 of particles carrying both electric and magnetic charge ({\bf dyons}).
Thus charges of both sign appear in the field theory associated to each
infrared (IR)  fixed point.

Moreover all the phases are subjected to the ``charge neutrality''
constraint, that is, the sum of the total electric and magnetic charge
carried by the condensed particles must be zero. This implies that, in each
phase, the number of ``particles'' must be equal to the number of
``antiparticles'', as usually happens within the condensate phases of a CG
\cite{nienhuis}.

In this paper we shall use a modified version of the mCG where an uniform
background charge is introduced \cite{noi}. In order to derive the phase
diagram we will perform the Renormalization Group ({\bf RG}) analysis of
the modified mCG. The relevant $SL(2, Z)$ (modular) symmetry analyzed in
\cite{car1,car2} is still a symmetry of the partition function, provided
one lets the transformations \underline{act onto the uniform background} as
well. That imposes strong constraints onto the global phase diagram, which
acquires the characteristic \underline{nested shape} \cite{scal2,scal1},
thus strongly suggesting an unified picture of the various  phases, as well
as of the
\underline{transitions}
 between phases. Then we are able to identify a
discrete infinite set of IR stable fixed points on the $\theta$-axis at
$\theta/2 \pi$ given by any rational number. Each of them controls the IR
physics of a region in the plane of the parameters $(   \frac{
\theta }{ 2 \pi} , \frac{ 1}{g} )$ and corresponds to a condensate of
dyons carrying both
electric and magnetic charge. The background charge and the particle
density do \underline{uniquely} fix the charges of the condensate.

The field theory that is supposed to describe the system close to an IR
fixed point breaks parity({\bf P}) and charge conjugation({\bf C}) (now the
charges of the dyons have all the same sign). It can be identified with a
2-dimensional({\bf 2D}) Conformal Field Theory ({\bf CFT}) with central
charge $c=1$, described by a chiral scalar field compactified on a circle
of radius $R^2
= q_e / q_m$, where $q_e$ and $q_m$ are the \underline{electric} and the
\underline{magnetic}
 charges of the dyons \cite{cmn}. That is a nontrivial result which, together with the
properties of the phase diagram, suggests the relevance of the model for
the 2D CFT description of the Fractional Quantum Hall Effect ({\bf FQHE}).

The paper is organized as follows:

\begin{itemize}

\item In section 2 we introduce the dimensionally reduced CR
model. Then we briefly sketch its \underline{gauge formulation} and focus
on its equivalent description as a $SL(2,Z)$ (modular) invariant mCG, which
is the one used for the RG analysis;

\item In section 3 we use a first-order RG calculation in order to derive
 the physical properties of the nonneutral phases, identified
by the condensation of dyons ;

\item In section 4 we perform  nonlinear RG calculations, which give
us the flow in the two parameters space ( $\frac{ \theta}{ 2
\pi}$,$\frac{1}{g}$),i.e. a complex plane;

\item In section 5 we get the global phase diagram by applying the
 RG results and the infinite discrete
 $SL(2, Z)$-symmetry, which acts on the complex plane, introduced in sec. 5.
This enables us to find all the RG stable fixed points and the relative
domains, where each condensate is relevant;

\item In section 6 we conjecture that the effective description at the
IR fixed points of our mCG is provided by a 2D chiral CFT. We briefly
explore the consequences of such an identification, particularly for what
concerns its relevance for the FQHE;

\item In section 7 we provide the main conclusions and discuss open problems.

\end{itemize}

\section{The CR model with a background charge density.}

In this section we shall construct the nontrivial mCG by properly
generalizing the model in \cite{car1,car2}. The starting point is the
mapping of the dimensionally reduced $Z_p$ model onto the mCG. The
introduction of the $\theta$-term allows for the mCG to have condensate
dyonic phases. The fundamental issue of charge neutrality \cite{nienhuis}
will be recovered by means of an uniform neutralizing background charge
density.

The original model is defined in terms of a two parameters lattice action:

\beq
S = \frac{1}{2 g} \sum ( \Delta_\mu \phi_j - S_{ \mu j} )^2 - i
\sum n_j \phi_j +
i \frac{ \theta }{ 4 \pi} \epsilon_{\mu \nu} \epsilon_{ij} \sum
( \Delta_\mu \phi_i - S_{ \mu i} ) ( \Delta_\nu \phi_j - S_{ \nu j } )
\label{action1}
\eneq

$\phi_j$, $j=3,4$, is a doublet of boson fields, $n_j$ is the ``electric
charge density'', the fields $S_{\mu j}$ are defined in terms of a constraint
that relates them to the ``magnetic charge distributions'' $m_j$:

\beq
\Delta_x S_{ y 4} - \Delta_y S_{ x 4 } = m_3 \;\;\; ;
\;
- \Delta_x S_{ y3} + \Delta_y S_{ x3} = m_4
\label{cons1}
\eneq

The corresponding ``Boltzmann weight'' in the total partition function is a
function of $n_j$ and $m_j$ and is given by:

\beq
\int \prod_j D \phi_j \int \prod_{\mu j}  D S_{\mu j}
\exp \left[ - S \right] \delta [ \Delta {\times} \vec{ S}_4 - m_3 ]
\delta [ \Delta {\times} \vec{S}_3 + m_4 ]
\label{cons2}
\eneq

Solving the constraint in eq.( \ref{cons2} ) one can either derive the ``gauge
representation'' ({\bf GR}) or the mCG representation of the CR-model.

In the GR one takes as dynamical variables the fields $\phi_j$ and the
Lagrange multipliers, $\tilde{\phi}_j$, introduced in order to solve
the constraint in eq.( \ref{cons2}) according to the identity:

\[
\delta ( \Delta_x S_{ y 4} - \Delta_y S_{ x 4} - m_3 ) =
\int D \tilde{ \phi}_3 \exp [ i \sum ( \Delta_x S_{ y 4} -
\Delta_y S_{ x 4} - m_3 ) ]
\]

 and a similar one ($4\leftrightarrow 3$) for the other constraint.

Integrating out the fields $S_{ \mu j}$ provides the action as a function of
the fields $\phi_j$ and $\tilde{ \phi}_j$. In the $\theta = 0$ case one
obtains:

\beq
S = \frac{g}{ 2} \sum ( \Delta_\mu \tilde{ \phi}_j )^2 + i \sum ( \vec{ \Delta}
\phi_4 {\times} \vec{ \Delta} \tilde{ \phi}_3 - \vec{ \Delta} \phi_3 {\times}
\vec{ \Delta} \tilde{ \phi}_4 ) + i \sum( n_j \phi_j + m_j \tilde{ \phi}_j )
\label{gr}
\eneq

Such a representation is useful in order to trace out the realization of
the mCG in terms of a 2D Conformal Field Theory as discussed, for example,
in \cite{noi}. Here we want to focus our analysis on the IR properties of
the system, which we shall analyze within the RG framework. The standard RG
analysis is usually carried out within the mCG representation
\cite{nienhuis}. Then from now on we shall use such a representation
throughout the rest of the paper. Here it is worth to mention that the
gauge invariance in the original 4-dimensional gauge theory \cite{car1} has
a remnant in the invariance of the action in eq.( \ref{gr}) under a
constant shift of the fields $\phi_j$, $\tilde{
\phi}_j$:

\[
\phi_j \rightarrow \phi_j + \alpha_j \;\; ; \;\;
\tilde{ \phi}_j \rightarrow \tilde{ \phi}_j + \beta_j
\]

The requirement of invariance under such a symmetry determines the charge
neutrality constraints:

\[
\sum n_j = \sum m_j = 0
\]

We will study solutions consistent with these constraints, where an uniform
background charge neutralizes the total charge of a condensate of
particles, all with the same charge.

At this point we must remark that, in performing the summations in eq.(
\ref{action1}), we have made no difference between the lattice and its
dual. We are alleged to do so because we use the lattice only in order to
regularize the gauge theory action. All the relevant RG analysis will be
carried out within the framework of the continuum CG model \cite{nienhuis}
and any distinction between the lattice and its dual will be lost anyway.

Let us now work out the mCG formulation with a background charge. The
$\theta$-term in eq.( \ref{action1})
 generates an additional ``minimal coupling'' that
adds up to the ``minimal electric coupling'' term to give:

\[
- i \sum \left( n_j + \frac{ \theta }{ 2 \pi} m_j \right) \phi_j
\equiv -i \sum q_j \phi_j
\]

Now we choose configurations where the charge densities are given by:

\beq
n_j ( r ) = \bar{n}_j + \nu_j ( r ) \;\;\; ; \;
m_j ( r ) = \bar{m}_j + \mu_j ( r )
\label{split1}
\eneq

where $\bar{n}_j, \bar{m}_j$ are uniform solutions so that the neutrality
conditions read as:

\[
\bar{ n}_j S + \sum_r \nu_j ( r ) \equiv \bar{N}_j + \sum_r \nu_j (r ) = 0
\]

\[
\bar{ m}_j S + \sum_r \mu_j ( r ) \equiv \bar{M}_j + \sum_r \mu_j ( r ) = 0
\]

where $S$ is the area of the sample.

According to the form of the charge density we have chosen, we can
split $S_{\mu j}$ as $S_{ \mu j} = \bar{S}_{\mu j} + \sigma_{ \mu j}$, where:

\[
\epsilon_{\mu \nu} \epsilon_{ ij} \Delta_\mu \bar{S}_{ \nu i } = m_j
\;\; \; ; \;
\Delta_\mu \bar{ S}_{ \mu j} = 0
\]

and

\beq
\epsilon_{\mu \nu} \epsilon_{ ij} \Delta_\mu \sigma_{ \nu i } = \mu_j
\label{split2}
\eneq

The solutions for $\bar{S}_{\mu j}$ are given by:

\beq
\bar{S}_{\mu j} = \epsilon_{\mu \nu} \Delta_\nu \epsilon_{ i j } \Phi_j ( r )
\eneq

where:

\[
\Phi_j ( r ) = \frac{ \bar{ m}_j}{4} r^2
\]

Moreover the total density of charge $q_j$ will be given by:

\beq
q_j = \bar{ n}_j + \frac{ \theta }{ 2 \pi} \bar{m}_j +
\nu_j + \frac{ \theta }{ 2 \pi} \mu_j \equiv
\bar{q}_j + \epsilon_j
\label{split3}
\eneq

Then we split also $\phi_j$ as follows:

\beq
\phi_j = \bar{ \phi}_j + \varphi_j
\label{split4}
\eneq

where

\[
\frac{1}{ g} ( \Delta)^2 \bar{ \phi}_j + i \bar{n}_j = 0
\]

The solutions for $\bar{\phi}_j$ are given by:

\[
i \bar{\phi}_j  = \frac{g \bar{n}_j}{4} r^2
\]

After disregarding irrelevant constant terms, one obtains the total action in
terms of the fluctuating fields and charge densities:

\beq
\tilde{S} = \frac{1}{2g} \sum ( \Delta_\mu \varphi_j - \sigma_{\mu j })^2
- i \sum \epsilon_j \varphi_j + \sum [ \Phi_j ( r ) \mu_j ( r ) +
\bar{ \phi}_j ( r ) \epsilon_j ( r ) ]
\label{action2}
\eneq

Eq.( \ref{action2} ) is almost the same result that one would obtain for a
neutral charge distribution \cite{car1, car2}. The only difference is
provided by the term:

\[
\sum [ \Phi_j ( r ) \mu_j ( r ) +
\bar{ \phi}_j ( r ) \epsilon_j ( r ) ]
\]

that describes the coupling of $\mu_j $ and $\nu_j$ to the background charge
densities.

Integration over the fields $\varphi_j$ is now straightforward. It yields:

\[
\tilde{S} =  - \sum_{ r , r^{'}}
\left[ \frac{ \mu_3 ( r ) \mu_3 ( r^{'} ) + \mu_4 ( r ) \mu_4 ( r^{'} ) }{ 2g}
 + \frac{ g}{2} [ \epsilon_3 ( r ) \epsilon_3 ( r^{'} ) +
\epsilon_4 ( r ) \epsilon_4 ( r^{'} ) ]
\right] G ( r ,  r^{'} ) +
\]

\beq
\sum [ \Phi_j ( r ) \mu_j ( r ) +
\bar{ \phi}_j ( r ) \epsilon_j
( r ) ] - i \sum_{ r r^{'} } ( \nu_3 ( r ) \mu_4
( r^{'} ) - \nu_4 ( r ) \mu_3 ( r^{'} ) ) \Theta ( r - r^{'} )
\label{gcg1}
\eneq

$G(r, r^{'} )$ is the longitudinal Green function of the fluctuating
field $\varphi_j$. It
must  be evaluated with a proper regularization. In the continuum limit it
is the usual Green function for a massless boson field in two dimensions:

\[
G(r, r^{'} )  \approx \ln \left| \frac{ r
- r^{'} }{ a } \right|
\]

where $a$ is a length scale that works as an UV regularizator of the theory.

The ``transverse'' Green function $\Theta ( r )$ is defined such that:

\beq
\partial_\mu \Theta ( r ) = \epsilon_{ \mu \nu} \partial_\nu G ( r, r^{'} )
\label{dual1}
\eneq

that implies:

\beq
\Theta ( r ) = {\rm arctan} \left( \frac{ y}{ x } \right)
\label{dual2}
\eneq

An important remark concerns the consequences of the Bohm-Aharonov ({\bf BA})
term that appears in eq.( \ref{gcg1}). It makes the partition function ill
defined, unless the combinations $\nu_i^3 \mu_j^4 - \nu_j^4 \mu_i^3$ are
integer numbers. From now on we will make the assumption
that $\nu_i^{3,4}$ and $\mu_i^{3,4}$ are integers  \cite{nienhuis}.

Eq.( \ref{gcg1} ) leads us to the first relevant result of our work. We
have written down a model for interacting particles with both electric and
magnetic charge. Each dyon takes either a type-3 or a type-4 charge. Then
we define the nonuniform part of the charge densities as a set of $N$
particles carrying either type-3 or type-4 charges:

\beq
\mu^j ( r ) = \sum_{k = 1}^N \mu_k^j \delta ( r - r_k ) \;\;\; ; \;
\nu^j ( r ) = \sum_{ k =1}^N \nu_k^j \delta ( r - r_k ) \; , j=3,4
\label{pochd}
\eneq

At fixed charge distribution the action will be given by \cite{noi}:

\[
S [ \{ \nu^j \} , \{ \mu^j \} ] =
\sum_{ i \neq k =1}^N \frac{1}{2} \left[ \frac{ \mu^3_i \mu^3_k +
\mu^4_i \mu^4_k}{g} +  g ( \epsilon_i^3 \epsilon_k^3 + \epsilon_i^4
\epsilon_k^4 ) \right] \ln \left| \frac{ r_i - r_k}{a} \right|
\]

\beq
- i \sum_{ i \neq k } ( \nu^3_i \mu_k^4 - \nu_k^4 \mu_i^3 ) \phi ( r_i - r_k)
- \frac{1}{ 4 } \sum_i \left( g ( \bar{q}^3 \epsilon_i^3 + \bar{q}^4
\epsilon_i^4 ) + \frac{ \bar{m} \mu^3_i + \bar{ m} \mu^4_i}{
g} \right) r_i^2
\label{action3}
\eneq

In order to perform a RG analysis of the mCG we need the grand canonical
partition function, $Z_{ GC} $. From now on we will neglect the index $j$
(we shall later justify such an assumption).
 $Z_{ GC} $ is written in terms of the ``fugacities'' $Y ( \nu , \mu )$,
given by the probability of creating a particle with charges $\nu , \mu$ at
fixed values of the parameters. In the continuum formalism
the mCG partition function is given by:

\beq
Z_{ GC} = \sum_{ \{ \nu , \mu \} } = \int \prod_j \left[ Y ( \nu_j , \mu_j )
\frac{ d^2 r_j}{a^2 } \right] e^{ - S [ \{ \nu , \mu \} ] }
\label{grand1}
\eneq

\section{Linear RG analysis.}

In this section we shall derive the RG equations at first order in the
fugacities and will use them in order to characterize the condensate phases.
The widely used RG analysis can be improved by
adding higher order (nonlinear) corrections in $Y$.

The (first-order) RG equations are derived by rescaling the cutoff,
$a \rightarrow a + d a$, and then by ``reabsorbing'' the corresponding
changes in the partition function by means of a redefinition of the
fugacities \cite{nienhuis}.

If $a \rightarrow a + da$, one has:

\[
 \int \prod_j \left[ Y ( \nu_j , \mu_j )
\frac{ d^2 r_j}{a^2 } \right] e^{ - S [ \{ \nu , \mu \} ] }
\rightarrow
\]

\beq
\int \prod_{ j = 1}^N \biggl\{
\left[ Y ( \nu_j , \mu_j ) \frac{ d^2 r_j}{a^2 } \right]
\left[ 1 - \frac{ da}{a} \left( 2 + \frac{ 1}{ 2 g} \mu_j ( \bar{M} -
\mu_j ) + \frac{ g}{2} \epsilon_j ( \bar{Q} - \epsilon_j )\right)
 \right] \biggr\}
e^{ - S [ \{ \nu , \mu \} ] } + O ( \left( \frac{ da}{ a} \right)^2 )
\label{rg1}
\eneq

In eq.(\ref{rg1}) $Y ( \nu_j , \mu_j )$ have to be understood as
``running'' fugacities, whose dependence on $a$ is fixed by the
requirement that $Z_{CG}$ \underline{does not depend on $a$}.
Accordingly, one gets the following renormalization of the fugacities:

\beq
Y ( \nu_j , \mu_j ) \rightarrow Y ( \nu_j , \mu_j ) + d Y ( \nu_j , \mu_j )
\label{rgf1}
\eneq

with

\[
a \frac{ d Y ( \nu_j , \mu_j )}{ d a }   =  x ( \nu_j , \mu_j )
Y ( \nu_j , \mu_j )
\]

\beq
x ( \nu_j , \mu_j ) = 2 + \frac{ 1}{ 2 g} \mu_j ( \bar{M} -
\mu_j ) + \frac{ g}{2} \epsilon_j ( \bar{Q} - \epsilon_j )
\label{rgf2}
\eneq

Following \cite{nienhuis} we assume that the set of particles that condense is
defined by the values of the charges that maximize each $x( \nu_j , \mu_j )$,
constrained by the global neutrality conditions:

\[
\sum_j \nu_j = - \bar{N} \;\;\;  ; \; \sum_j \mu_j = - \bar{M}
\]

The solution is given by:

\beq
\nu_1 = \ldots = \nu_N \equiv \bar{ \nu} = - \frac{ \bar{N}}{N} \;\;\; ; \;
\mu_1 = \ldots = \mu_N \equiv \bar{ \mu} = - \frac{ \bar{M}}{ N}
\label{nneut1}
\eneq

and the corresponding value of the exponent is:

\beq
x ( \bar{ \nu} , \bar{ \mu} ) = 2 + \frac{ 1}{ 2g} \bar{\mu} ( \bar{M} -
\bar{\mu} ) + \frac{ g}{ 2 } \bar{q} ( \bar{Q} - \bar{q} )
\approx 2 + \frac{ 1}{ 2 g} \bar{\mu} \bar{M} + \frac{g}{2} \bar{q} \bar{Q}
\label{nneut2}
\eneq

where we have introduced the ``generalized charges''
$\bar{q} = \bar{\nu} + \frac{ \theta }{ 2 \pi } \bar{\mu} $.

The outcome of the linearized RG analysis in the presence of a background
is that the condensate phases are given by a gas of particles carrying both
electric and magnetic charges ({\bf dyons}) of the same sign and all
identical, corresponding to a breaking of both P and C. Being all the
charges equal, the BA term disappears and $S [
\{ \mu^j \} , \{ \nu^j \} ]$ becomes the sum of two independent
contributions. This justifies our neglecting the index $j$ and the BA-term
in the RG analysis.

The breaking of the discrete symmetries P and C implies that the underlying
field theory breaks them as well, and we believe it to be related  to what
happens in a chiral 2D CFT \cite{cmn}.

The partition function for the condensate of $N$ dyons with charges $\bar{
\mu}$, $\bar{ \nu}$ is given by:

\beq
Z^*_{ \bar{ \mu} \bar{ \nu}} =
\int \frac{ d^2 r_1}{ a^2} \ldots \frac{ d^2 r_N}{ a^2} \exp \left[ \left(
\frac{ 1}{ 2 g } \bar{\mu}^2 + \frac{ g}{2} \bar{ \epsilon}^2 \right)
\sum_{ i \neq j } \ln \left| \frac{ r_i - r_j}{ a} \right| -
\frac{ 1}{4} \left( \frac{ \bar{ \mu} \bar{m} }{ g} + g \bar{ q}
\bar{ \epsilon} \right) \sum_j r_j^2 \right]
\label{partf}
\eneq

\section{Nonlinear RG approximation.}

In this section we shall extend the analysis of section[3] by including
higher order processes that eventually lead to a renormalization of the
interaction strength between dyons.

In order to write down higher order RG equations we will perturb around the
condensate with a fluctuation given by a particle/antiparticle pair. This
is just a different way of describing the ``charge annihilation'' process
analyzed in \cite{nienhuis}. The action for the condensate plus the pair is
given by:

\[
S = S_N + \Delta S
\]

where $S_N$ is the action for the unperturbed condensate and:

\beq
\Delta S = \left( \frac{ \mu_P^2}{ 2 g} + \frac{ g}{2} \epsilon_P^2 \right)
\ln \left| \frac{ R_+ - R_-}{ a } \right| - \sum_{ j =1}^N \left( \frac{
\mu_P \bar{\mu}}{ 2 g } + \frac{ g}{2} \epsilon_P \bar{q} \right)
\biggl\{ \ln \left| \frac{ R_+ - r_j}{a} \right| - \ln \left| \frac{ R_- -
r_j }{ a } \right| \biggr\}
\label{pert}
\eneq

Notice that we have neglected the BA-term; infact we shall prove that
$\mu_P,\nu_P$ must be equal to $\bar{\mu},\bar{\nu}$, which justifies that.
$R_+$ and $R_-$ are the coordinates of the particle and of the
antiparticle, respectively, $\pm \mu_P $ and $\pm \epsilon_P$ are the
charges of the two particles in the pair.

In order to find the correction to the coupling constant, we must sum over
all configurations with $a < | R_+ - R_- | < a + d a$. Following
\cite{nienhuis}, we define:

\[
R_+ = R + \frac{s}{2} \;\;\; ; \; R_- = R - \frac{ s}{2}
\]

At second order in $s$ we have:

\beq
\Delta S = \left( \frac{ \mu_P^2}{ 2 g} + \frac{ g}{2} \epsilon_P^2 \right)
\ln \left| \frac{ s}{ a} \right| + \frac{1}{2} \left[ \frac{ \mu_P \bar{m}}{
g } + \frac{ g}{ 2 }\epsilon_P
\bar{n} \right] s {\cdot} R - \sum_j \left( \frac{
\bar{ \mu} \mu_P}{ 2 g } + \frac{ g}{2} \bar{q} \epsilon_P \right) s {\cdot}
\frac{ (R - r_j )}{ | R - r_j |^2}
\label{pert2}
\eneq

Summing over the relevant range of values of $s$, we get the following
contribution to the $N$-particle condensate partition function:

\beq
\int \frac{ d^2 R}{ a^2 } \int_{ a < |s| < a + da } \frac{ d^2 s}{a^2 }
|s|^{ -  \left( \frac{ \mu_P^2}{ 2 g} + \frac{ g}{2} \epsilon_P^2 \right) }
\exp \biggl\{ s {\cdot} \frac{ \partial}{ \partial R} \psi ( R ) \biggr\}
 Y ( \mu_P , \nu_P ) Y ( - \mu_P , - \nu_P )
\label{pert3}
\eneq

where the function $\psi$ is given by:

\[
\psi (R ) = - \left( \frac{ \bar{ \mu} \mu_P}{ 2 g} + \frac{ g}{2} \bar{ q}
\epsilon_P \right) \sum_j \ln \left| \frac{ R - r_j}{ a} \right| +
\frac{1}{4} \left[ \frac{ \mu_P \bar{m}}{ 2 g} + \frac{ g}{2}
\epsilon_P \bar{q} \right] R^2
\]

Let us define $y^2_{\mu_P , \nu_P }
  \equiv Y ( \mu_P , \nu ) Y ( - \mu_P , - \nu_P )$. The linear RG equation
for $y_{\nu_P , \nu_P}^2$ is:

\beq
\frac{ d y_{ \mu_P , \nu_P}^2}{ d \ln a} = 2 \left[ 2 - \frac{ \mu_P^2}{ 2g}
- \frac{ g}{2} \nu_P^2 \right] y_{ \mu_P , \nu_P}^2
\label{renfp}
\eneq

Then the operator $y_{ \mu_P , \nu_P}$ is relevant for

\beq
2 - \frac{ \mu_P^2}{ 2 g} -\frac{ g}{2} \left( \nu_P + \frac{ \theta}{  \pi}
\mu_P \right) > 0
\label{rel1}
\eneq

Eq.( \ref{rel1}), together with the $SL(2, Z)$-symmetry of the model (see
next section), allows us to determine the global phase diagram. For example
we shall see
 that shifting $\frac{ \theta}{2 \pi}$ by -1 is equivalent to
shifting $\nu_P$ to $\nu_P + \mu_P$ and $\bar{N}$ to $\bar{ N} + \bar{M}$.
The important consequence of the discrete symmetry is the map between
regions of parameter space corresponding to different condensate phases,
one onto the other. This generates the global phase diagram that is quite
similar to the one derived in \cite{car2} (see Fig.1). However the
background
\underline{ does definitely} modify the RG flow, as we will see later.

Eq.( \ref{pert3} ) may be expanded at second order in $s$. The result is:

\[
{\rm const} - \frac{ \pi}{2} y^2 \frac{ da }{ a} \int d^2 R \psi ( r )
\nabla^2 \psi ( r )
=
\]

\beq
{\rm const} - \frac{ \pi}{2} y^2 \frac{ da}{a} \left( \frac{ \mu_P \bar{
\mu}}{ 2g} + \frac{ g}{2} \epsilon_P \bar{ q} \right) \sum_{ i \neq j = 1}^N
\ln \left| \frac{ r_i - r_j}{a} \right| + \frac{ \pi}{2} y^2 \frac{ 1}{2}
\frac{ da}{a} \left( \frac{ \mu_P \bar{m}}{ 2 g } + \frac{ g}{2} \nu_P
\bar{n} \right) \sum_{ j = 1}^N r_j^2
\label{pert4}
\eneq

Eq.( \ref{pert4} ) corresponds to a renormalization in the coupling
constant:

\[
\alpha^2_{ \bar{\mu} \bar{ \nu }} = \frac{ \bar{ \mu}^2 }{ g^2 } + g \bar{q}^2
\]

In fact, by summing over all the possible fluctuations, one obtains:

\beq
d \alpha^2_{ \bar{\mu} \bar{ \nu }} = - \sum_{ \nu_P , \mu_P} \frac{ \pi}{2}
y_{ \mu_P , \nu_P }^2 \frac{ da}{a} \left( \frac{ \mu_P \bar{ \mu}}{ 2 g }
+ \frac{ g}{2} \epsilon_P \bar{q} \right)^2
\label{pert5}
\eneq

With our choice of the parameters only one of the $y_{ \mu_P  \nu_P}^2$ is
relevant. Consistency with the linear RG analysis shows that the relevant
one is $y_{ \bar{ \mu} \bar{\nu}}$. Then we can approximate the right hand
side of eq.( \ref{pert4}) as:

\beq
\frac{ d \alpha_{ \bar{\mu} \bar{\nu}} }{ d \ln a} = - \frac{ \pi}{2}
y^2_{ \bar{ \mu} \bar{ \nu}} ( \alpha_{ \bar{ \mu} \bar{\nu} })^2
\label{pert6}
\eneq

 The solution of eq.( \ref{pert6})is given by:

\beq
\frac{ 1}{ \alpha_{ \bar{ \mu} , \bar{ \nu}}( a )} -
\frac{ 1}{ \alpha_{ \bar{ \mu} , \bar{ \nu}}( a_0 )} =
\frac{ \pi}{ 2} \int_{ a_0}^a \frac{ da}{a} y_{ \bar{ \mu} , \bar{ \nu}}^2
\label{pert7}
\eneq

At very large scales we find the values of the interaction strength
corresponding to the IR fixed point of the system. The fixed point strength
is given by:

\beq
\lim_{ a \rightarrow \infty } \alpha_{ \bar{ \mu} , \bar{ \nu}} ( a )
=0
\label{pert8}
\eneq

Since $\alpha_{ \bar{ \mu } , \bar{ \nu}}$ is a nonnegative definite
function, eq.( \ref{pert8} ) is satisfied if, and only if:

\beq
\lim_{ a \rightarrow \infty } \frac{ 1}{ g  ( a ) } = 0 \;\;\; ; \;
\lim_{ a \rightarrow \infty } \frac{ \theta }{ 2 \pi } ( a ) = -
\frac{ \bar{ \nu}}{ \bar{ \mu}}
\label{ pert9}
\eneq

Eq.( \ref{ pert9} ) tells us what the IR fixed points of the model are in
the plane ($\frac{ \theta}{ 2
\pi}$ , $\frac{ 1}{ g}$), precisely they all lie on
the real axis at $\frac{\theta}{2 \pi }$ equal to any rational number. Such
a result is deeply related to the properties of the group $SL( 2 , Z)$.
Infact all the RG fixed points of the mCG are $SL( 2 , Z)$-fixed points
too. Then it is possible to generate the manifold of all these points by
letting $SL(2, Z)$ act on anyone of them, and consequently the local RG
flow is mapped also onto the flow around any other fixed point.

\section{ Discrete symmetries and global phase diagram.}

In the previous sections we have employed RG equations in order to find
informations about the condensate phases of the system. The RG approach
only provides local informations on the phase diagram. Instead its global
properties are a consequence of the discrete group $SL(2, Z)$(the modular
group), which is a symmetry of the model.

Introducing the complex variable $\zeta
=
\frac{
\theta }{ 2 \pi } + i \frac{ 1}{ g}$,
 the group $SL( 2, Z)$ acts as:

\beq
\zeta \rightarrow \frac{ A \zeta + B}{ C \zeta + D}
\label{sl2z1}
\eneq

where:

\[
M \equiv \left( \begin{array}{ cc} A & B \\ C & D \end{array} \right)
\]

is a matrix whose entries are relative integers and whose determinant is
one. It is well known \cite{gin} that the full group is generated by
repeated application of the following two transformations:

\begin{itemize}

\item {\bf Translations} \[ T \;\; :  \zeta \rightarrow \zeta + 1 ; \]

\item {\bf Inversions} \[ S \;\; :  \zeta \rightarrow -1/\zeta \]

\end{itemize}

The action of $SL(2,Z)$ on the mCG is defined both on the parameter space
and on the charge distributions as follows:

\[
\zeta \rightarrow \zeta + 1 \;\; ; \; \mu \rightarrow \mu \;\; ; \; \nu
\rightarrow \nu + \mu \;\; ; \; \bar{M} \rightarrow \bar{M} \;\; ; \;
\bar{N} \rightarrow \bar{N} + \bar{M}
\]

\beq
\zeta \rightarrow - \frac{ 1}{ \zeta} \;\; ; \; \mu \rightarrow - \nu
\;\; ; \; \nu \rightarrow \mu \;\; ; \; \bar{ M} \rightarrow - \bar{N} \;\; ;
\; \bar{N} \rightarrow \bar{M}
\label{mappe}
\eneq

i.e., they act simultaneously on the complex parameter space and on the
background, which fixes also the dyon charges.

Nonlinear RG analysis provides us with the relevance condition for $y_{ \bar{
\mu} \bar{ \nu}}^2$, eq.(\ref{rel1} ).
Such a condition determines a domain in the $\left( \frac{ \theta}{ 2 \pi},
\frac{ 1}{g } \right)$-plane we shall refer to as the ``relevance domain'' for
the  dyon condensate of charges $\bar{ \mu} , \bar{ \nu}$. It is
appropriate to notice that the dyonic phases with $ \frac{ \theta}{ 2\pi}
\neq 0$ realize in 2D the t'Hooft confined phases \cite{gerard}.

Using the full $SL(2, Z)$-symmetry, as defined in eq.(\ref{mappe}), one can
generate the regions of attraction for all possible condensates, which do
not overlap. The phase boundaries are ``semicircle'' mapped one onto the
other by $SL( 2 , Z)$-transformations. Therefore we have uniquely
determined the phase diagram and its boundaries, which cover all the upper
right complex plane (see Fig.1). We should notice that the modular group
should also help in deriving the full beta-function in whole the upper
complex plane in terms of modular forms (see \cite{tan} for an attempt).

\section{The critical theory for the fixed points: relation to FQHE.}

In this section we make clear our main conjecture about the field theory
description of the attractive fixed points.

To derive the critical theory from the general CG model is a difficult
task. Still one can try to guess it and check its consistency with the
general description in terms of a mCG. In the following we propose such a
field theory, at the IR fixed points, as the 2D CFT description \cite{cmn}
of the QH condensates at Laughlin's fillings and can eventually be
generalized at least for Jain's fillings \cite{cmm}.

As it has been seen, all the phases break $P$ and $C$ and they correspond
to a Coulomb Gas with an uniform classical background. The regions of
attraction are determined by the value of the background charge densities
and are related one to the other by $SL(2,Z)$-transformations. This
determines a ``hierarchy'', so that the relevance domains``shrink'' as one
goes deeper into the hierarchy level (see Fig.1). Finally the charge ratios
for the condensate are in agreement with the interpretation in terms of
quasiparticles with fractional charge \cite{bob}. Then, we hypothesize
that, at the IR stable fixed points, the correct field theory limit of our
model is given by a chiral CFT with central charge $c=1$, described in
terms of a compactified scalar field with radius given by $R^2 = 2 p +1$.
In such a theory, the conformal blocks correspond to the vertex operators:

\[
: e^{ i \alpha_l \phi ( z )} : \;\; ; \; \alpha_l =
\frac{ l}{ R} ; l = 1\ldots 2p+1
\]

The momentum lattice $\{ p /R\}$ and the weight lattice $\{ R^2 w \}$
describe the eletric and the magnetic charge of ``anyons'', respectively.
When $l=2p+1$ we get the ``electron'' operator. Such a theory has been
successfully employed in the description of the FQHE plateaux at filling
$\nu
= 1/(2p+1)$ \cite{cmn}. In conclusion it seems
 appropriate to us to identify the field-theoretical description of the IR fixed
point with $\frac{\theta}{2\pi} = - \frac{
\bar{\nu}}{\bar{\mu}}$ with one of the CFT described above at the proper
value of $R$.

By acting with $SL(2,Z)$ on a single IR fixed point one generates all the
others, thus strengthening the idea of an unified description of all the IR
fixed points (and the Hall plateaus we identify with them). Previous work
on this subject \cite{noi,carp} seems to strongly support such an
identification, too.

On the other hand the mGG, here presented, is directly connected with
Laughlin plasma description \cite{bob}. Infact it is well kwown that the
square modulus of the ground state wavefunction of a QH-condensate at
filling $1/(2p+1)$ is given by:

\beq
| \Psi ( z_1 , \ldots , z_{N_e} )|^2 = \prod_{ i < j =1}^{ N_e} | z_i - z_j
|^{2( 2 p+1) } e^{ - \frac{1}{ 2} \sum_{j=1}^{ N_e} | z_j|^2}
\label{bobo}
\eneq

 It is evident that eq.(\ref{bobo}) is exactly the partition function in eq.(\ref{partf}) for
a dyonic condensate whose electric-to-magnetic charge ratio, given by
$-\frac{ \theta_*}{ 2 \pi}$, is equal to $1/(2 p +1)$.

So we can be confident on the correctness  of such a guess, which should be
thoroughly studied.

\section{Conclusion and perspectives.}

In this paper we have discussed a modified version of the CR mCG. The
presence of the background term in such a generalized CG model allows for
some new properties:

\begin{enumerate}

\item Getting non-neutral condensate phases, related by $SL ( 2 , Z)$ as
its attractive fixed points. They are rational points on the
$\frac{\theta}{2\pi}$-axis (see Fig.1).  The related condensate is made out
of dyons with electric charge $\bar{\nu}$ and magnetic charge $\bar{\mu}$,
such that $\frac{ \theta_*}{ 2 \pi} = - \frac{ \bar{\nu}}{
\bar{\mu}}$.

\item Driving the RG flow of the system by means of tunable external
parameters, given by the electric (magnetic) background densities
$\bar{N}$($\bar{M})$;

\item Defining the global phase diagram by means of the $SL( 2 ,Z)$-symmetry;

\end{enumerate}

A possible field theory for the attractive fixed points has been proposed
as a chiral CFT, at least for the simple case $\frac{\theta_*}{2
\pi} =
\frac{1}{ 2 p + 1}$. Then the mCG seems to be consistent with the
physics of the FQHE. In this framework the identification of the parameters
$\frac{1}{ g}$ and $\frac{\theta}{2 \pi}$ as the longitudinal and the Hall
conductance, respectively, appears natural. However the denominator of
$\frac{ \theta_*}{ 2 \pi}$, $\bar{\mu}$, is not constrained to be odd, as
it should be. This is related to the wrong choice of the discrete symmetry
group, the right one being the group $\Gamma (2)$ discussed in
\cite{lr1,lr2}. Anyway the relation of the present model with the FQHE
physics deserves further work. We look forward to completing the analysis
of the global phase diagram in particular of the unstable fixed points,
which describe the transition between two different fillings. The modular
invariance should imply, in our opinion, the (super) universality of the
transition, advocated by experimental \cite{sondhi} and theoretical work
[laws of corresponding states \cite{klz}]. As a final remark we point out
the possible role of such a model in the study of non-commutative gauge
theories.

\vspace{1.5truecm}

\centerline{\bf Acknowledgments}

Work supported in part by the European Commission RTN programme
HPRN-CT-2000-00131.

\newpage

\begin{figure}[t]
\begin{center}
\epsfig{file=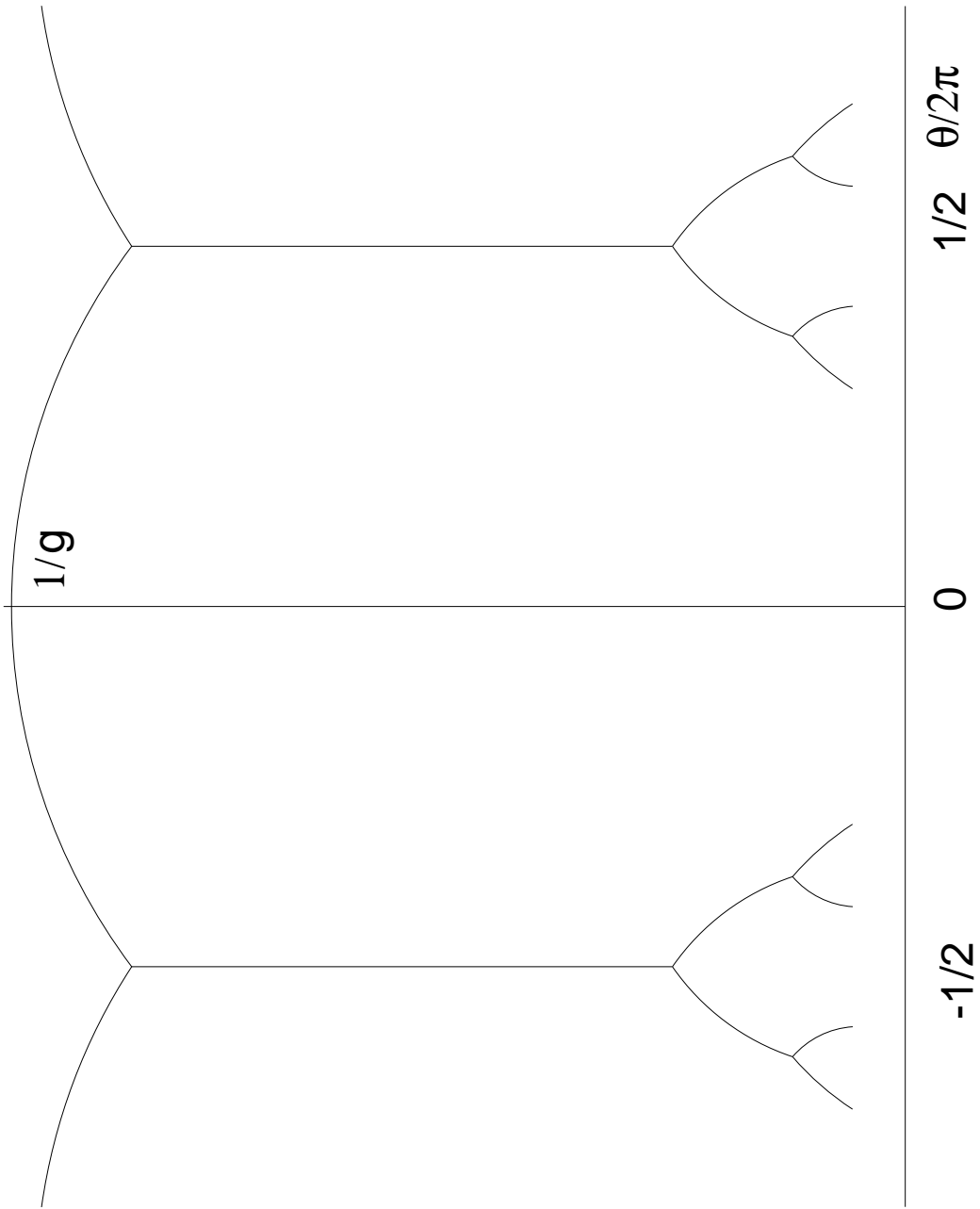,height=17truecm}
\end{center}
\caption{The global structure of the phase diagram}
\end{figure}

\newpage

\begin{thebibliography}{99}

\bibitem{gerard} G. 't Hooft, Nuc. Phys. {\bf B 138}(1978)1.


\bibitem{car1} J.L. Cardy and E. Rabinovici, Nucl. Phys. {\bf
B 205}(1982)1.

\bibitem{car2} J.L. Cardy, Nucl. Phys. {\bf B 205}(1982)17



\bibitem{nienhuis} B.Nienhuis,
"Coulomb Gas Formulation of Two-dimensional Phase
Transitions",C. Domb and J.Lebowitz Eds.,
(Academic Press) (1987).

\bibitem{noi} G. Cristofano, D. Giuliano and G. Maiella,
J. Phys. I France {\bf 7} (1997)1033-1038.


\bibitem{scal2} A. M. M. Pruisken, Phys. Rev. Lett. {\bf 61
N.11}(1988)1297.

\bibitem{scal1} H. P. Wei, D. C. Tsui, M. A. Paalanen, A. M. M.
Pruisken, Phys. Rev. Lett. {\bf 61 N. 11}(1988)1294.

\bibitem{cmn} G. Cristofano, G. Maiella, R. Musto and F. Nicodemi,
Nuc. Phys. {\bf B}(Proc. Suppl.){\bf 33 C} (1993)119.

\bibitem{gin} P. Ginsparg, ``Applied Conformal Field Theories", les
Houches lectures {\bf Vol.49}, E. Brezin and J. Zinn-Justin eds.(1988).

\bibitem{lr1} C.A. L\"{u}tken and G.G. Ross, Phys. Rev.{\bf B 48}(1993)2500

\bibitem{lr2} C.A. L\"{u}tken and G.G. Ross, Phys. Rev.{\bf B 45}(1992)11837

\bibitem{carp} D. Carpentier, Jour. Phys. {\bf A 32} no.21(1999)3865.

\bibitem{bob} R.B. Laughlin, "Elementary Theory of Incompressible Quantum
Fluid" in "The Quantum Hall Effect", R,E. Prange and S.M. Girvin
Eds(Springer,1987).

\bibitem{cmm} G. Cristofano, G. Maiella and V. Marotta, Mod. Phys. Lett.
{\bf A15} (2000)547.

\bibitem{tan} N. Taniguchi, cond-mat/9810334.

\bibitem{sondhi} E. Shimshoni, S.L. Sondhi and D. Shahar, Phys. Rev. {\bf B 55}
(1997)13730.

\bibitem{klz} S. Kivelson, D.H. Lee and S.C. Zhang,Phys. Rev.{\bf B46}
(1992)2223

\end{thebibliography}
\end{document}